\documentclass[sigconf]{acmart}
\usepackage{multirow}
\usepackage{subcaption}
\usepackage{algorithm}
\usepackage{algpseudocode}

\AtBeginDocument{%
  \providecommand\BibTeX{{%
    \normalfont B\kern-0.5em{\scshape i\kern-0.25em b}\kern-0.8em\TeX}}}

\begin{document}

\title{FedChain: An Efficient and Secure Consensus Protocol based on Proof of Useful Federated Learning for Blockchain}


\author{Peiran Wang}
\affiliation{%
  \institution{Institute of Network Sciences and Cyberspace, Tsinghua University}
  \city{Beijing}
  \country{China}}
\email{wangpr22@mails.tsinghua.edu.cn}



\renewcommand{\shortauthors}{Trovato and Tobin, et al.}

\begin{abstract}
Blockchain has become a popular decentralized paradigm for various applications in the zero-trust environment. The core of the blockchain is the consensus protocol, which establishes consensus among all the participants. PoW (Proof-of-Work) is one of the most popular consensus protocols. However, the PoW consensus protocol which incentives the participants to use their computing power to solve a meaningless hash puzzle is continuously questioned as energy-wasting. To address these issues, we propose an efficient and secure consensus protocol based on proof of useful federated learning for blockchain (called FedChain). We first propose a secure and robust blockchain architecture that takes federated learning tasks as proof of work. Then a pool aggregation mechanism is integrated to improve the efficiency of the FedChain architecture. To protect model parameter privacy for each participant within a mining pool, a secret sharing-based ring-all reduce architecture is designed. We also introduce a data distribution-based federated learning model optimization algorithm to improve the model performance of FedChain. At last, a zero-knowledge proof-based federated learning model verification is introduced to preserve the privacy of federated learning participants while proving the model performance of federated learning participants. Our approach has been tested and validated through extensive experiments, demonstrating its performance.
\end{abstract}




\maketitle

\section{Introduction}
The decentralized blockchain paradigm has emerged as a prominent technology that enables secure and transparent distributed transactions without relying on a central authority. Blockchain can be viewed as a decentralized distributed ledger that is shared among all participating nodes in the network. The consensus protocol, which plays a pivotal role in ensuring the security and integrity of the system, is typically the central component of the blockchain. The consensus protocol requires all participating nodes to agree on a common ledger without the presence of a centralized entity, thus ensuring consistency across all the ledgers in the blockchain. Proof-of-Work (PoW), one of the most widely used consensus protocols, incentivizes each node to compete for the ownership of the next ledger by solving a computationally challenging but meaningless cryptographic puzzle. The first node to solve the puzzle broadcasts the solution to all other nodes in the network, which then verify the solution and update their local blockchain ledgers with the new information. This process, commonly referred to as "mining," completes the consensus, and the miner who solves the puzzle first is rewarded with the ownership of the next ledge. Blockchain has been applied in various domains \cite{li2021blockchain, nguyen2021federated, pokhrel2020federated, lu2020blockchain}, such as supply chain management, healthcare, and finance, due to its inherent features of immutability, transparency, and decentralization. Moreover, several blockchain-based consensus protocols, such as Proof-of-Stake (PoS), Delegated Proof-of-Stake (DPoS), and Byzantine Fault Tolerance (BFT), have been developed to address the scalability and energy consumption issues of the PoW protocol.

\par The traditional PoW consensus protocol, while effective at establishing consensus among miners, has raised concerns regarding the significant energy consumption associated with solving meaningless hash puzzles. Reports indicate that the annual energy consumption of the Bitcoin blockchain could power a country such as Thailand for an entire year. Moreover, the PoW protocol does not contribute to any useful tasks beyond securing the blockchain.

\par To address these concerns, Proof-of-Useful-Work (PoUW) has emerged as a promising alternative, where miners perform useful tasks, such as searching prime chains \cite{king2013primecoin}, image segmentation \cite{li2019exploiting}, or all-pairs shortest path \cite{ball2017proofs}, instead of solving meaningless hash puzzles. Recently, researchers have explored the possibility of introducing machine learning tasks as useful work as the consensus protocol \cite{chenli2019energy, qu2021proof, wang2022platform}. Deep learning models, in particular, have a high demand for computing power and energy and are easy to verify, making them an attractive option for use as useful work. However, training a large AI model requires a significant amount of training data, which is typically not owned by a single miner. Additionally, acquiring data from other miners could compromise their privacy, making centralized deep learning infeasible.

\par To address this issue, researchers have introduced federated learning into the blockchain as PoUW. Federated learning is a distributed machine learning paradigm that trains a global model without direct access to each participant's private data. In federated learning, a central server initially generates a global model and then sends it to each participant at the beginning of each training round. Each participant then updates the global model using their local private data and uploads weight updates to the central server. The central server aggregates all the weight updates to generate a new global model. This approach allows the central server to train the global model using weight updates from each participant, instead of their private data.

\begin{table}[t]
  \caption{Summary of Notations}
  \resizebox{0.5\textwidth}{!}{
  \begin{tabular}{lp{6cm}}
    \bottomrule
    Term&Description\\
    \hline
    $n$ & the node in the blockchain \\ 
    $N$ & the nodes in the blockchain \\ 
    $p$ & the mining pool in the blockchain \\
    $t_p$ & estimated time for $p$ to complete model training \\
    $P$ & the mining pools in the blockchain \\
    $m$ & the model trained by $p$ \\
    $l_{i,j}$ & the communication latency between $n_i$ and $n_j$ \\
    \hline
  \end{tabular}
  }
\end{table}

Overall, the introduction of federated learning-based PoUW into the blockchain represents a significant step forward in mitigating energy consumption, while simultaneously enabling the development of innovative blockchain applications in the field of machine learning. However, previous works PoFL \cite{qu2021proof} and PF-PoFL \cite{wang2022platform} have made significant progress in introducing federated learning as proof of work, however, there remain unconsidered problems. 
\par \textbf{1)} These approaches do not consider the potential security threats posed by attackers, leaving them vulnerable to attacks; 
\par \textbf{2)} Previous approaches adopt centralized federated learning, which violates the principles of blockchain, thus making their architecture untrustworthy and unstable;
\par \textbf{3)} Previous researchers also do not provide efficient private protection mechanisms for the participants of federated learning within the blockchain consensus protocol which will harm the anonymity of the blockchain (other participants can infer each participant's identity from its federated learning model update);
\par \textbf{4)} To prove the accuracy of the model trained by federated learning, previous researchers use homomorphic encryption as the basis for model accuracy verification, which is a computationally-intensive task and could significantly increase the computational cost of the consensus establishment process. 
\par To address these challenges, we propose an efficient secure consensus protocol called FedChain, which takes federated learning as proof of useful work for blockchain. The FedChain architecture incorporates a pool aggregation mechanism to enhance its efficiency rather than taking all the clients participating in a single federated learning procedure. To safeguard the model parameter privacy of each participating node within a mining pool, we have designed a secret sharing-based ring-all reduce architecture. Our approach includes a data distribution-based federated learning model optimization technique, which boosts the model's performance in FedChain. Lastly, we introduce a zero-knowledge proof-based federated learning model verification method to ensure the privacy of participants while verifying the model's performance. In summary, FedChain offers a more efficient, secure, and trustworthy architecture for using federated learning as proof of work in blockchain-based applications, while also maintaining the privacy of the participants and protecting the system against potential security threats. In FedChain, we made the following contributions:

\begin{itemize}
    \item We propose FedChain, a secure and robust blockchain architecture that utilizes federated learning as proof of work. A new block structure and transaction types are introduced under FedChain to enhance its security and robustness. A decentralized federated learning design is also presented to counter the complex environments of Byzantine attackers and unstable centralized nodes, which is more effective than directly applying the traditional centralized paradigm.
    \item Considering the high cost brought by the traditional federated learning process participated by all the nodes, we introduce a pool aggregation mechanism to split the participating nodes of the blockchain into different mining pools to accelerate the consensus establishment process.
    \item To protect model parameter privacy within a mining pool, a secret sharing-based ring-all reduce architecture is designed. Each participant adds noise to its parameter part and performs ring-all reduction on it.
    \item To address the non-iid data distribution in FedChain's federated learning settings, KL divergence is used to describe the data distribution and is employed as an aggregation weight to improve the model optimization process.
    \item Finally, the zero-knowledge proof is introduced to verify each model's accuracy without obtaining each mining pool's model before reward settlement.
\end{itemize}

\section{Related Work}
Recent studies have examined energy-recycling consensus techniques, where the energy expended on difficult but pointless puzzles in proof-of-work is recycled to carry out real-world helpful tasks, called proof of useful work (PoUW). Ball et al. \cite{ball2017proofs} provide proof of useful work based on a variety of computational issues, such as Orthogonal Vectors, 3SUM, All-Pairs Shortest Path, and any problem that reduces to them (this includes deciding any graph property that is statable in first-order logic). Proof of Learning (PoLe), a novel consensus protocol proposed by Liu et al. \cite{liu2021zkcnn}, uses block consensus computation to optimize neural networks. In their approach, the consensus nodes train NN models on the training and testing data, which acts as proof of learning, and then disclose the results to the entire blockchain network. They create a secure mapping layer (SML), which can be easily implemented as a layer of linear NN, as a fundamental part of PoLe to stop consensus nodes from lying. A new block is added to the blockchain after the network's consensus has been reached. Li et al. \cite{li2019exploiting} provide a system that allows miners to conduct image segmentation as PoUW rather than calculating pointless hash values, allowing them to take advantage of the computing power of blockchain miners for biomedical image segmentation.

\section{System Model}
In this section, we discuss the system model and our design goals.
\begin{figure*}[htb!]
\centering
\includegraphics[width=0.9\textwidth]{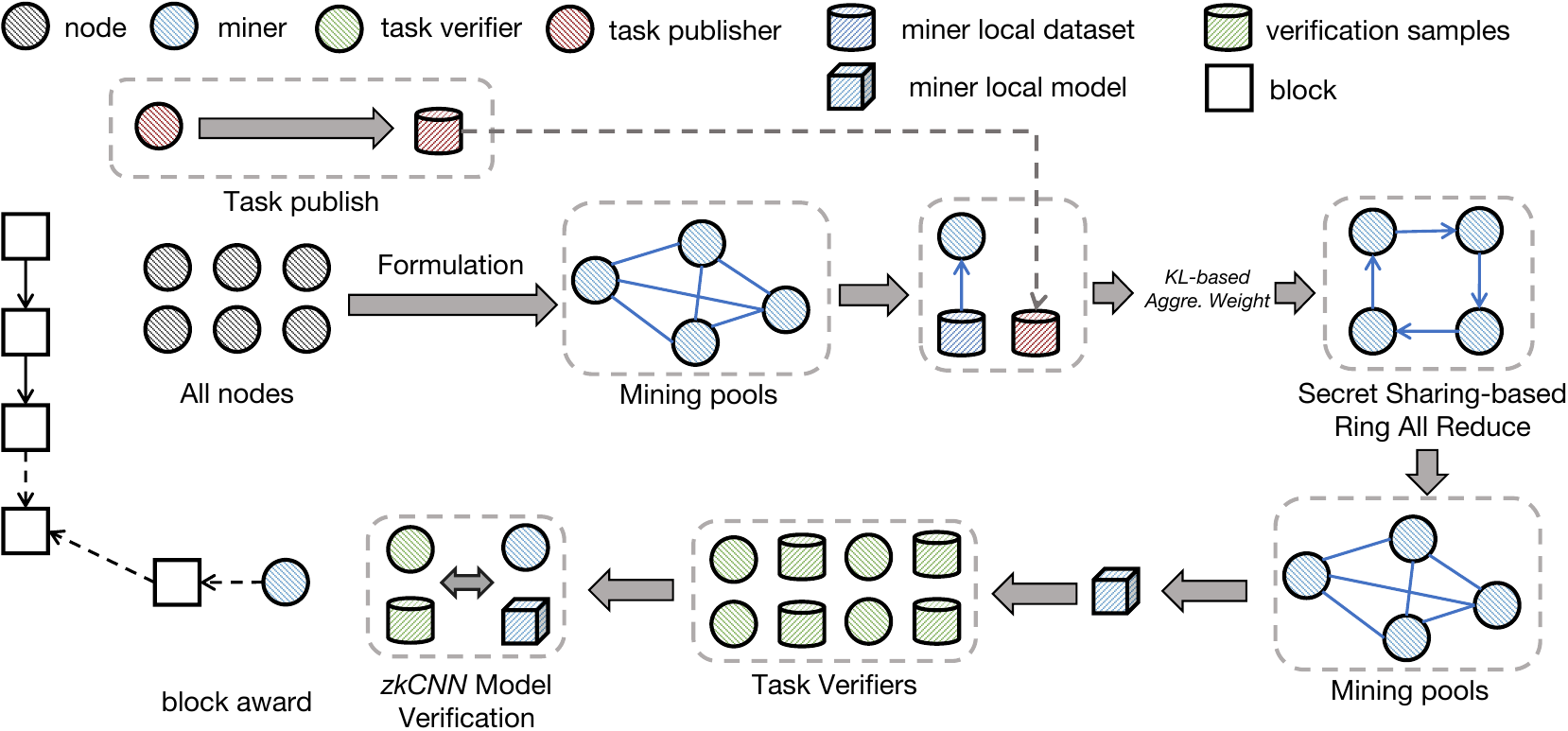}
\caption{The architecture of the FedChain framework.}
\label{fig:framework}
\end{figure*}
\subsection{Blockchain System Model}
We first show the system architecture of FedChain with three entities, task publisher, miner, and model verifier. Each node can simultaneously perform the three roles.
\begin{itemize}
    \item \textbf{Task Publisher:} Each blockchain node has the ability to take on the role of a task publisher, publishing FL tasks (such as semantic analysis and biomedical image recognition) to the blockchain platform with clear completion deadlines, an example dataset $d$, performance standards, and rewards as incentives. 
    \item \textbf{Miner:} Each blockchain node has the ability to mine independently or as part of a mining pool. The mining pool will then use federated learning to have all of its miners do the tasks that task publishers have advertised. 
    \item \textbf{Task Verifier:} Each blockchain node has the ability to act as a task verifier, who is in charge of confirming the model that the miners have provided. The model verifier should determine whether miners' claims that their model can achieve a given level of accuracy are accurate.
\end{itemize}
\subsection{Design Goals}
\label{sec:design_goals}
The key objective of FedChain is to provide an efficient, private, and trustworthy decentralized federated learning scheme as the proof of useful work for consensus protocol in the blockchain. Our design goals are given as follows:
 \begin{itemize}
     \item Traditional federated learning will randomly pick some nodes to participate or allow all the nodes to participate. However, in the real world, it will not be efficient since some participants may have bad network status. To make the federated learning executes efficiently, we introduced a mining pool aggregation process that allows the nodes in the blockchain to unify as different mining pools to execute in-pool federated learning.
     \item Existing work mainly directly applies centralized federated learning on the blockchain which violates the zero-trust hypothesis of blockchain.  We designed a decentralized federated learning architecture that enables the federated learning to execute without a central server. Furthermore, to protect each participant's model parameter privacy, we design a secret sharing-based ring all-reduce scheme.
     \item The federated learning within the mining pool faces the non-iid data distribution which may result in low performance of the global model. Thus, we design a data distribution-based model optimization technique.
     \item How to validate the accuracy of the global model provided by nodes has been a critical problem. Existing work focuses on using an encryption-based method like homomorphic encryption to validate node models' accuracy while maintaining their privacy. We apply zero-knowledge proof-based model validation to validate the models provided by each mining pool.
 \end{itemize}

\section{Proposed Scheme}
In this section, we show the proposed FedChain consensus protocol. Derived from the design goals in \ref{sec:design_goals}, we propose the mining pool aggregation, secret sharing-based ring all reduce decentralized federated learning, data distribution-based federated learning optimization, and zero knowledge proof-based model verification into the design of FedChain consensus protocol.

\subsection{Mining Pool Aggregation}

The high latency between the nodes in our blockchain will bring high communication costs to our FedChain consensus protocol, thus making FedChain unusable. To mitigate the latency problems brought by federated learning, we design a mining pool aggregation algorithm. We formulate this problem as a clustering problem. It consists of 4 steps:

\par \textbf{Step-1-Latency Estimation:} In round $\tau$, each node $n_i$ can compute the estimated latency $l'_{i,j}$ between itself and other node $n_j$ in the node set $N$ derived from its history communication latency. The history of communication latency between $n_i$ and $n_j$ is
\begin{equation}
    l^1_{i,j}, l^2_{i,j}, ..., l^{\tau-1}_{i,j}
\end{equation}
Then $n_i$ can compute the estimated latency:
\begin{equation}
    l'_{i,j}=\frac{\sum_{x}^{\tau-1}l^x_{i,j}}{\tau-1}
\end{equation}

\par \textbf{Step-2-Head Node Announcement:} Then some nodes $H={h_1, h_2, ..., h_{|P|}}$: can declare themselves as the head node for the mining pool:
\begin{equation}
    H={h_1, h_2, ..., h_{|P|}}.
\end{equation}

\par \textbf{Step-3-Node Mining Pool Choose:} Then the remaining nodes $N$ in the node set that does not announce themselves as head nodes will choose which mining pool $p$ they will join. The join process is guided by the estimated latency. 
Each remaining node $n$ will compute the estimated time cost if $n$ join the mining pool $p={n_{p,1}, n_{p,2}, ..., n_{p,|p|}}$ as
\begin{equation}
    l'_{n,p}=max(t_{p}, max(l'_{n,n_{p,1}}, l'_{n,n_{p,2}}, ..., l'_{n,n_{p,|p|}})).
\end{equation}

\subsection{Secret Sharing-based Ring All Reduce Decentralized Federated Learning}

\begin{figure*}[htb!]
\centering
\includegraphics[width=\textwidth]{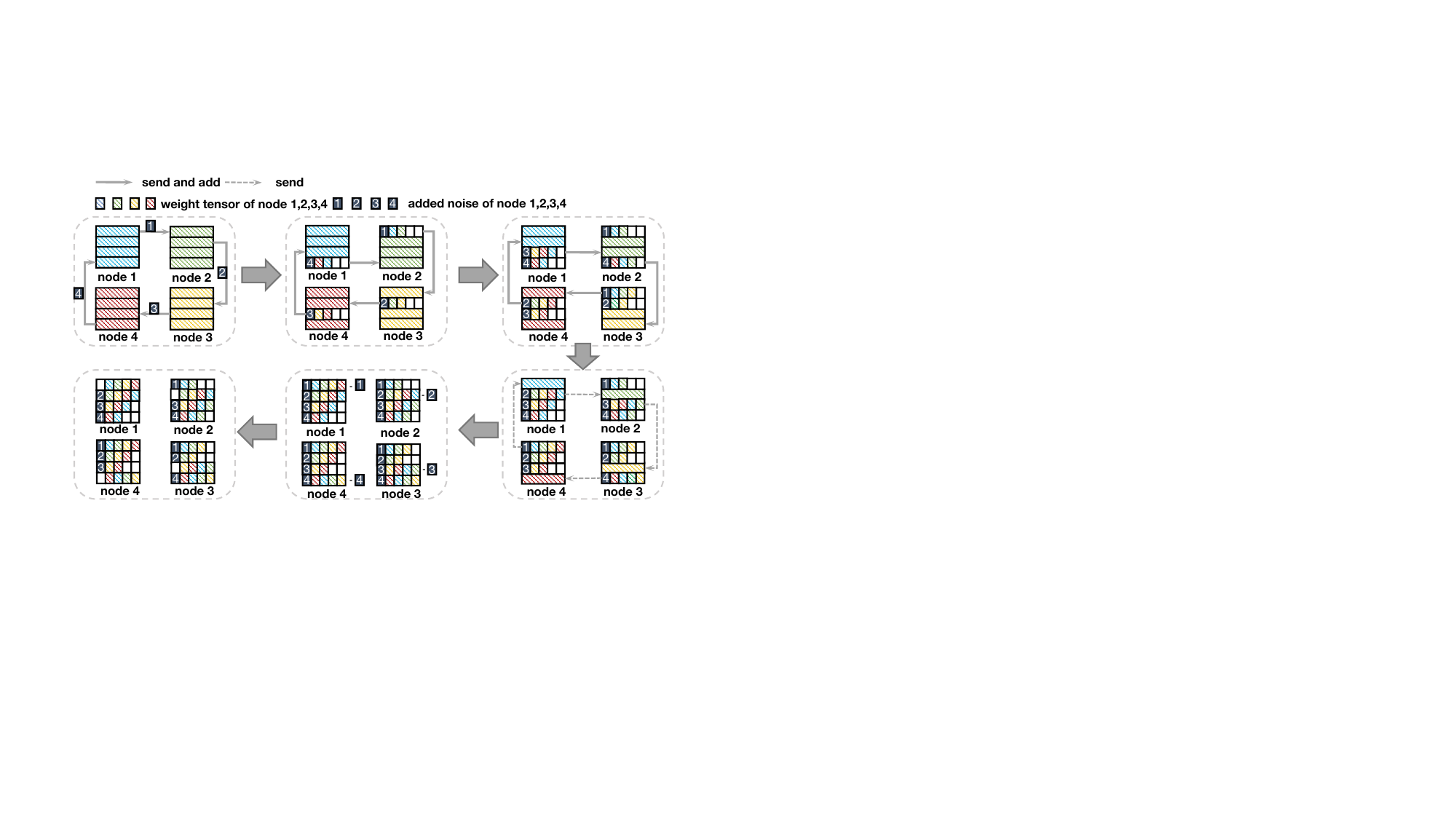}
\caption{Secret Sharing Ring All Reduce.}
\label{fig:privateRingAllReduce}
\end{figure*}

After the formulation of mining pools, each mining pool $p$ needs to train its deep learning model $m_p$. The initial federated learning design mainly applies centralized federated learning. However, the untrusted central server (malicious or unstable central server) will cause the centralized federated learning to be invalid. Furthermore, applying a centralized application in the zero-trust context of blockchain is unacceptable as well. Therefore, we introduce the secret sharing ring all reduce-based decentralized federated learning architecture:

\par Each miner $n_i$ within the mining pool $p$ will train a local model
\begin{equation}
    m_i=TrainModel(n_i).
\end{equation}
Then, $n_i$ will split the model into $|p|$ pieces according to the number of miners within $p$
\begin{equation}
    m_i\to \begin{vmatrix}
            \omega_{i,1} \\  \omega_{i,2} \\ ... \\ \omega_{i,|p|-1} \\ \omega_{i,|p|}
        \end{vmatrix}.
\end{equation}
$n_i$ will generate noise $b_i$ whose size is similar to the size of $\omega_{i,i}$, and translates the splited model as 
\begin{equation}
    \begin{vmatrix}
        \omega_{i,1} 
        \\  \omega_{i,2} 
        \\ ... 
        \\ \omega_{i,i}+b_i
        \\ ... 
        \\ \omega_{i,|p|-1} 
        \\ \omega_{i,|p|}
    \end{vmatrix}.
\end{equation}
Then the miners will perform ring all reduce as shown in Fig. \ref{fig:privateRingAllReduce}. The miner $n_i$ will get
\begin{equation}
    \omega'_{i,i}=b_i+\sum_{i}^{|p|} \omega_{i,i}.
\end{equation}
Then it can perform 
\begin{equation}
    \omega'_{i,i}=\omega'_{i,i}-b_i
\end{equation}
to get the complete $\sum_{i}^{|p|} \omega_{i, i}$ without leaking any other node's private model parameter within the ring all reduce process.

\begin{algorithm}[t]
\caption{Secret Sharing-based Ring All Reduce} 
\label{alg:system}
\hspace*{0.02in} {\bf Input:} 
a mining pool $p$ with the miners $\left \{ n|n\in p \right \} $ within it\\
\begin{algorithmic}[1]
    \For{$n_i|n\in p$}
    \State $m_i=TrainModel(n_i)$
    \State $m_i\to \left | \omega_{i,1},...,\omega_{i,|p|}  \right | $
    \State $b_i=GenerateNoise(n_i)$
        \For{$j=1,2,3,...,|p|$}
        \State $\omega'_{i,j}=\omega_{i,j}$
        \EndFor   
    \State $\omega'_{i,i}=\omega_{i,i}+b_i$
        \For{$r=0,1,2,...,|p|-1$}
        \State $\omega'_{i,(i-r)\%||p|}=\omega'_{i-1,(i-1-r)\%||p|}+\omega'_{i,(i-r)\%||p|}$
        \EndFor
    \State $\omega'_{i,i}=\omega'_{i-1,i}-b_i$
    \EndFor
    \Return $\left \{ \omega'_{i,i}|i=1,2,...,|p| \right \} $
\end{algorithmic}
\end{algorithm}

\subsection{Data Distribution-based Federated Learning Optimization}

Assume a mining pool $p$ with some miners $\left \{ n_i | n_i \in p \right \}$ within it. Each miner $n_i$ has a local dataset $d_i$ and dataset size $|d_i|$. So the total size for $p$ is $D_p=\sum^{|p|}_{i}|d_i|$. The global loss function is $F(m)$. And the loss funcion for miner $n_i$ on dataset $d_i$ is
\begin{equation}
    f_i(m_i)=\ell(m_i, d_i)=\sum^{|d_i|}_{j}\ell(m_i,x_{i,j},y_{i,j}).
\end{equation}
The goal of machine learning is to find
\begin{equation}
    m_{i,opt}=argmin f_i(m_i)=argmin \sum^{|d_i|}_{j}\ell(m_i,x_{i,j},y_{i,j})
\end{equation}
thus getting the optimal model.

\par Furthermore, the model update process in each training process $\tau$ can be expressed as 
\begin{equation}
    m_i^{\tau+1}=m_{i}^{\tau}-\partial f_i(m_i).
\end{equation}
Due to the global loss function of federated learning
\begin{equation}
    F(m)=\sum^{|p|}_{i}\frac{|d_i|}{|D_p|}f_i(m_i) 
\end{equation}
the loss function can be further expressed as
\begin{equation}
    \partial F(m)=\sum^{|p|}_{i}\frac{|d_i|}{|D_p|}\partial f_i(m_i).
\end{equation}
The above all is the FedAvg algorithm which directly takes each participant's dataset size as its aggregation weight. However, in the real-world setting where non-iid data distribution affects the performance of the model deeply, directly taking dataset size as the weight is not a good choice. Therefore, we design a data distribution-based model optimization process.

\par Different datasets have different data distribution features. To describe the data distribution of a given dataset, we introduce the \textit{KL-divergence}. The data distribution between 2 datasets can be described as 
\begin{equation}
    D_{KL}(d_{i}||d_{j})=\sum d_{i}(X)log_{2}\frac{d_{i}(X)}{d_{j}(X)}
\end{equation}
$X$ is the classification type while $d_{i}(X)$ indicates the frequency of label $X$ within $d_{i}$. The intuitive thinking is that the smaller $n_i$'s dataset $d_i$'s KL divergence to the dataset example $d$ given by task publisher $T$, the better its model performs on it. So the data distribution-based loss function can be derived as
\begin{equation}
    \partial F(m)=\sum^{|p|}_{i}\sum (1-d_{i}(X)log_{2}\frac{d_{i}(X)}{d(X)})\partial f_i(m_i).
\end{equation}

\subsection{Zero Knowledge Proof-based Task Verification}

\par In this section, we introduce a zero-knowledge proof-based federated learning task model accuracy verification. After the federated learning training process, the nodes $\left \{ n_i | n_i \in p \right \}$ within each mining pool $p$ will have a model $m_p$. $\left \{ n_i | n_i \in p \right \}$ need to prove to other nodes who perform as model verifiers that $m_p$ can achieve a certain accuracy on a given dataset $X$ while not leaking the detailed parameter of $m_p$. If the miners directly send their models to the task verifiers, the detailed model parameter of their models will be leaked out. To solve this issue, we introduce a zero-knowledge proof scheme $zkCNN$ \cite{liu2021zkcnn}.

\par $zkCNN$ is a zero-knowledge proof scheme for machine learning model owners to prove to others that the owners' models do achieve certain accuracy on a given dataset using the models. While in the process of $zkCNN$, the detailed parameters of the models do not need to be sent to the others from the model owners thus protecting the model's privacy. The detailed steps of $zkCNN$ are shown below.
\par Let $X$ be a data sample. Let $y=pred(m_p,X)$ be the prediction of $X$ using the model. The $zkCNN$ scheme consists of the following steps:
\begin{itemize}
    \item $pp \gets  zkCNN.KeyGen(1^\lambda )$: Given the security parameter $1^\lambda$, task verifiers generate the public parameters $pp$ using the $zkCNN$. Then the verifiers will send $pp$ to the miners.
    \item $com_{m_p} \gets zkCNN.Commit(m_p, pp, r)$: The miners will use $zkCNN.Commit$ to encrypt the private model $m_p$ using $pp$ and $r$. The generated model commit $com_{m_p}$ will be sent to all the task verifiers.
    \item $(y, \pi ) \gets zkCNN.Prove(m_p, X, pp, r)$: The task verifiers give a data sample $X$, then the miners use the algorithm to run the CNN prediction algorithm to get $y = pred(m_p, X)$ and generate the proof $\pi$. Then the miners will send the $(y, \pi)$ to the task verifiers.
    \item ${0, 1} \gets zkCNN.Verify(com_{m_p}, X, y, \pi, pp)$: The task verifiers use the algorithm to verify the prediction $y$ with the commitment $com_{m_p}$, the proof $\pi$ and the input $X$.
\end{itemize}

\par $zkCNN$ scheme is sound, where the probability that $p$ returns a wrong prediction and passes the verification is negligible; it is also zero knowledge, where the proof leaks no information about the mining pool $p$’s model $m_p$. In the federated learning model verification step, each miner will repeat the $zkCNN$ predictions on multiple data samples and compares the predictions with the labels to calculate the accuracy. The introduction of $zkCNN$ into our scheme can help participants to preserve the privacy of miners while proving the model performance of miners. Thus miners do not need to reveal their model details to other nodes while proving the validity of their models.


\begin{figure*}[!htp]
  
  \centering
    \subfloat[100 nodes \& 5 mining pools]{\includegraphics[width=0.25\textwidth]{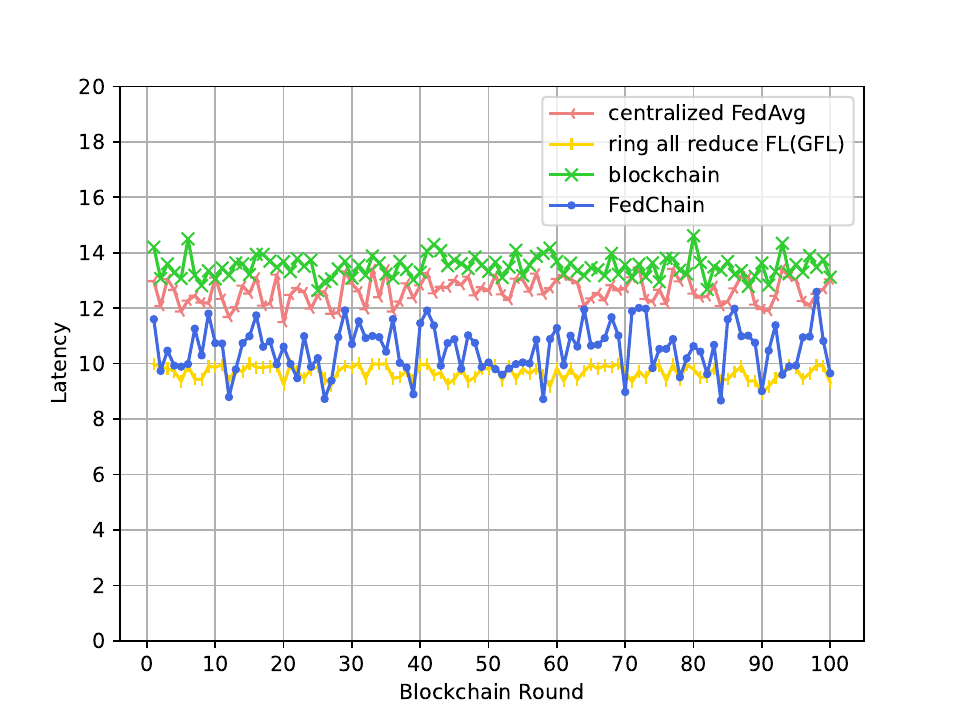}}
    \subfloat[100 nodes \& 10 mining pools]{\includegraphics[width=0.25\textwidth]{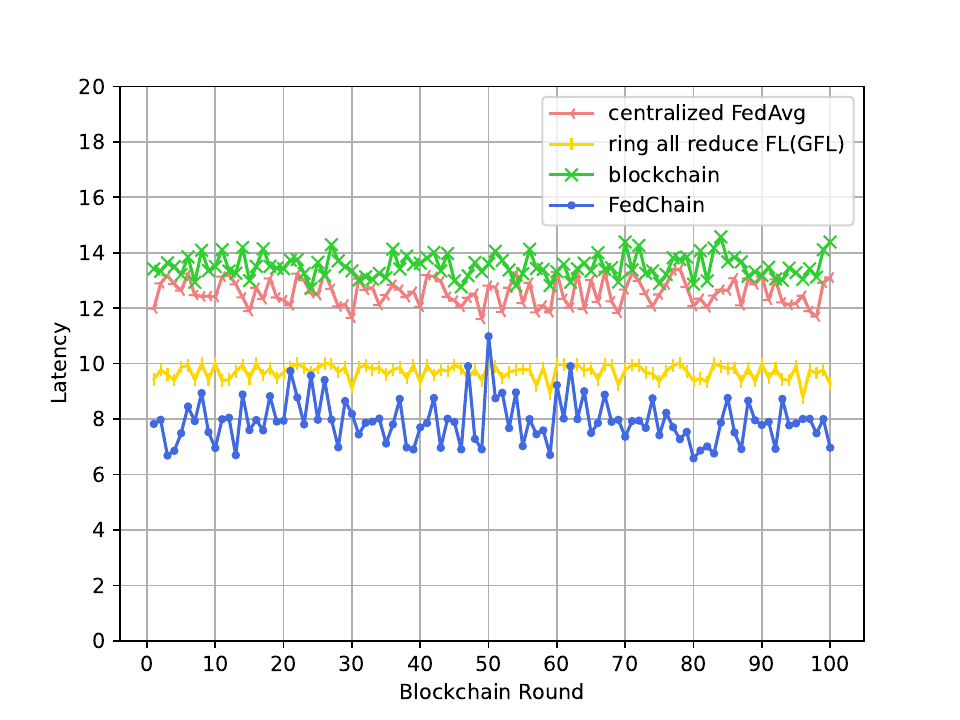}}
    \subfloat[100 nodes \& 20 mining pools]{\includegraphics[width=0.25\textwidth]{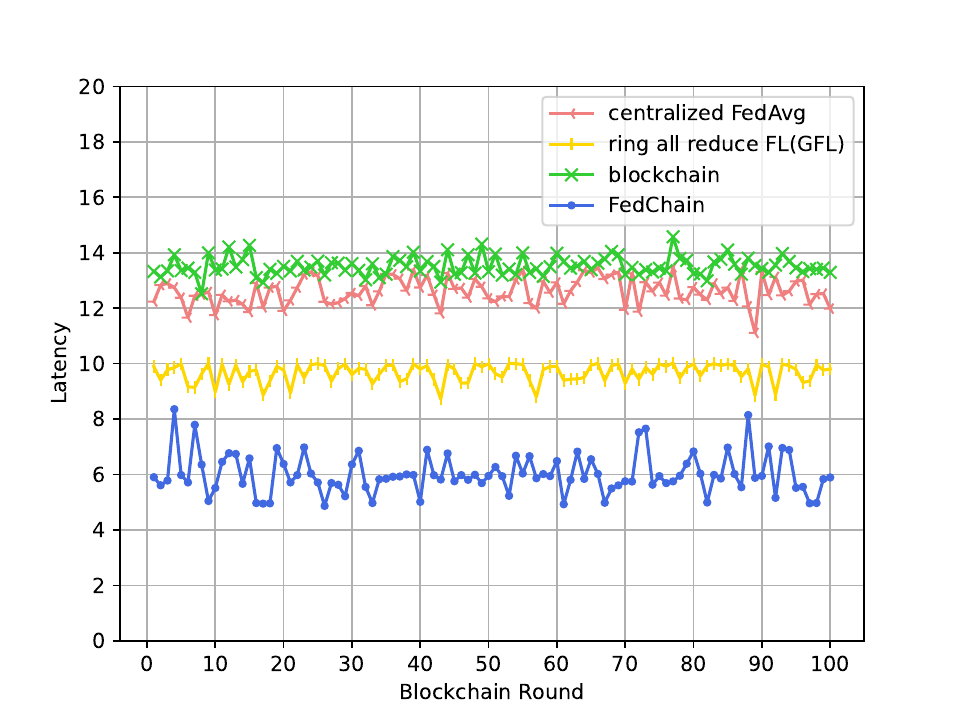}}
    \subfloat[100 nodes \& 50 mining pools]{\includegraphics[width=0.25\textwidth]{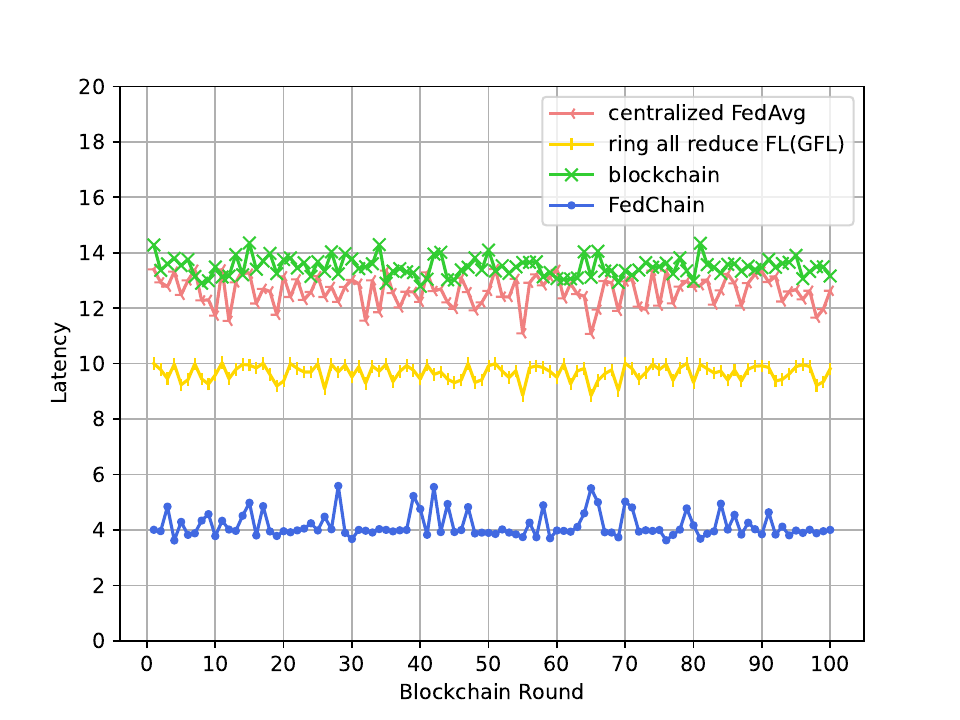}}
    \\
    
    \subfloat[200 nodes \& 5 mining pools]{\includegraphics[width=0.25\textwidth]{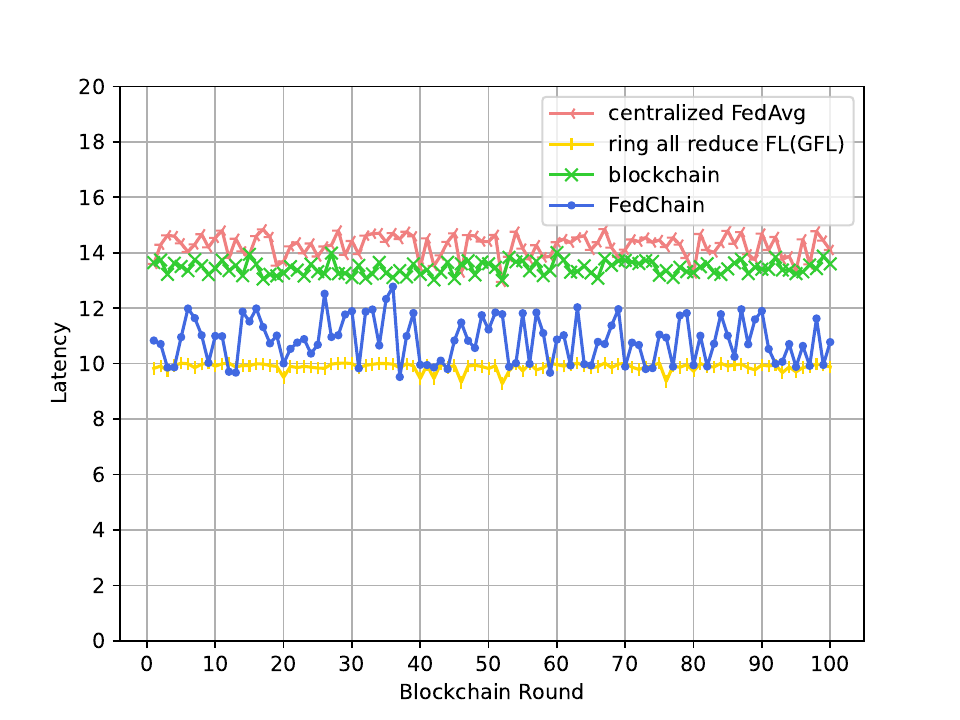}}
    \subfloat[200 nodes \& 10 mining pools]{\includegraphics[width=0.25\textwidth]{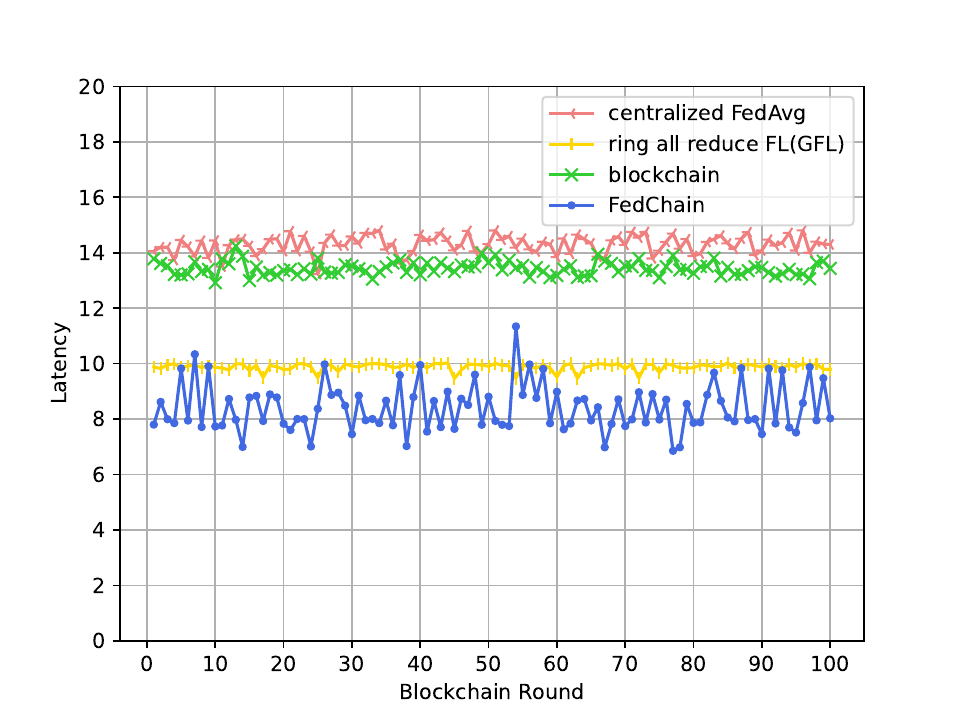}}
    \subfloat[200 nodes \& 20 mining pools]{\includegraphics[width=0.25\textwidth]{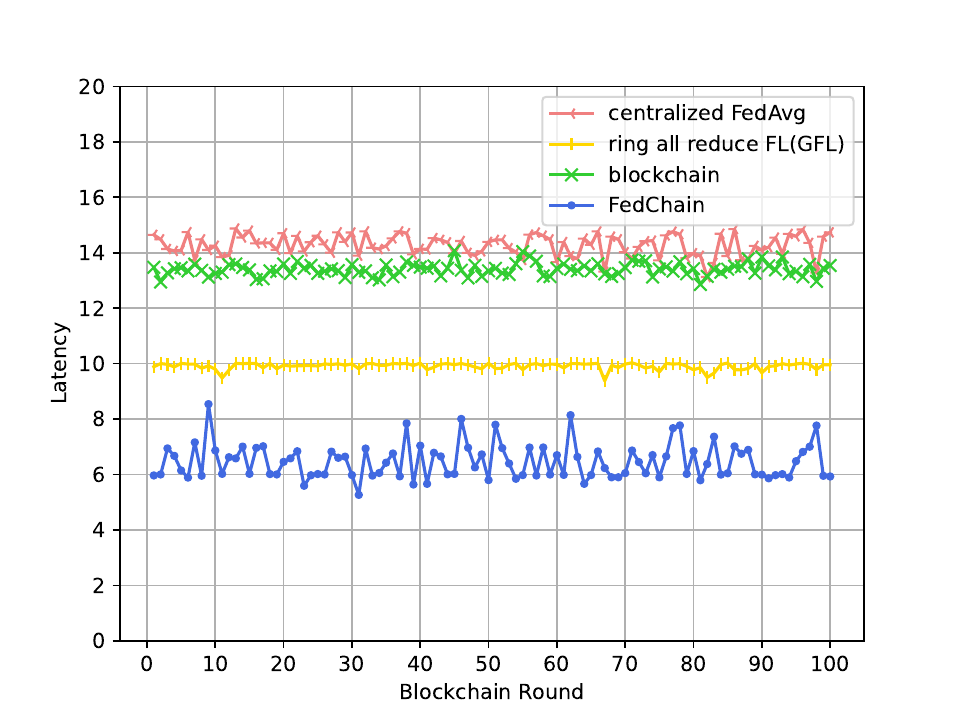}}
    \subfloat[200 nodes \& 50 mining pools]{\includegraphics[width=0.25\textwidth]{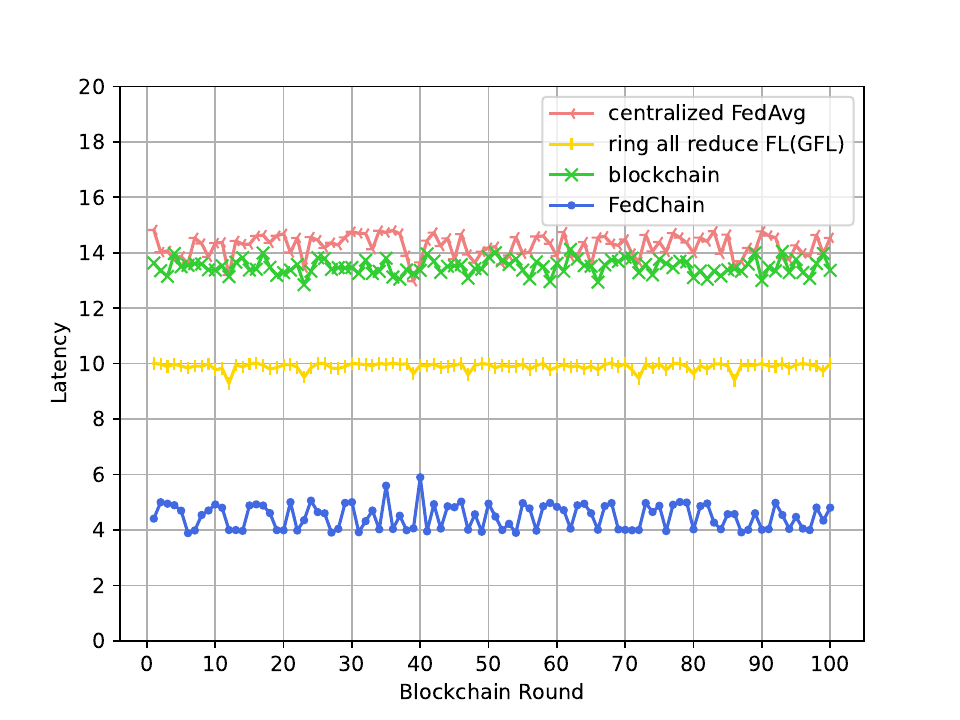}}
  \\
  
    \subfloat[500 nodes \& 5 mining pools]{\includegraphics[width=0.25\textwidth]{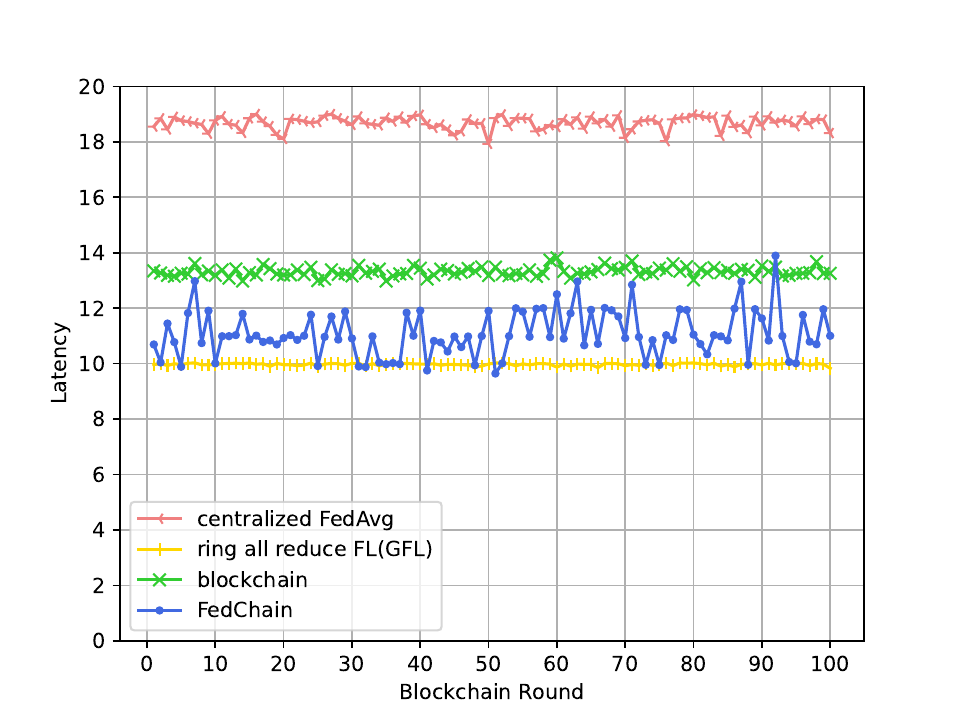}}
    \subfloat[500 nodes \& 10 mining pools]{\includegraphics[width=0.25\textwidth]{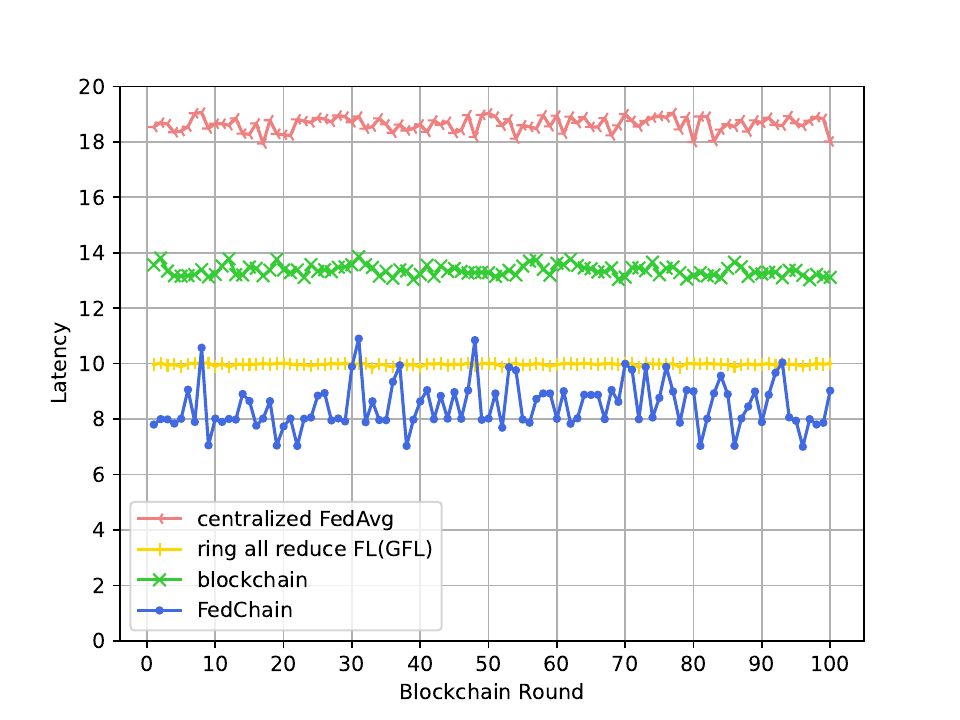}}
    \subfloat[500 nodes \& 20 mining pools]{\includegraphics[width=0.25\textwidth]{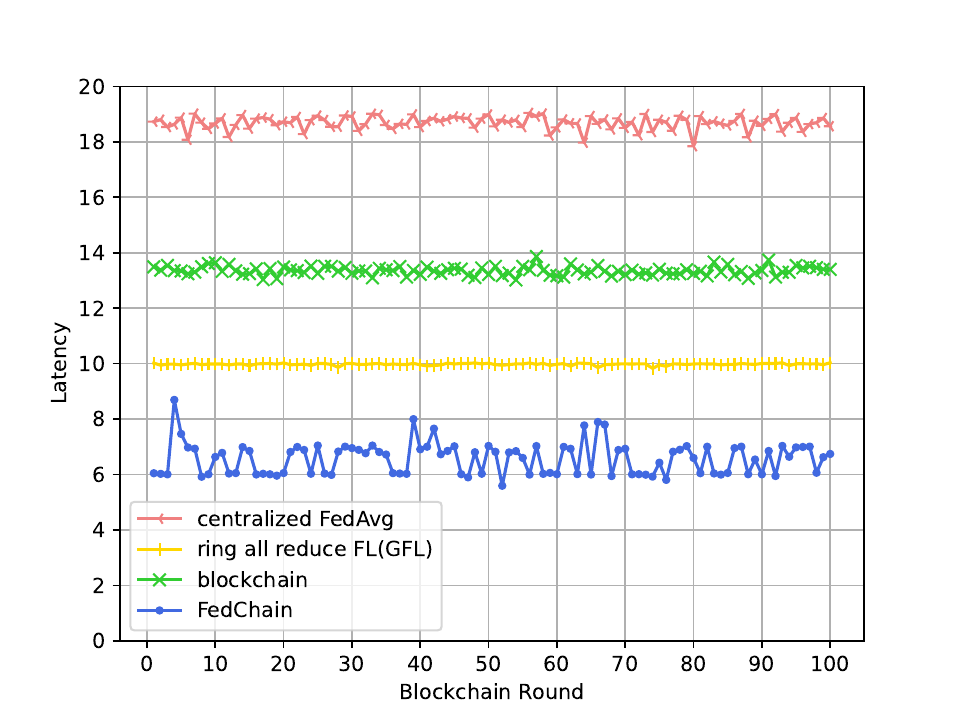}}
    \subfloat[500 nodes \& 50 mining pools]{\includegraphics[width=0.25\textwidth]{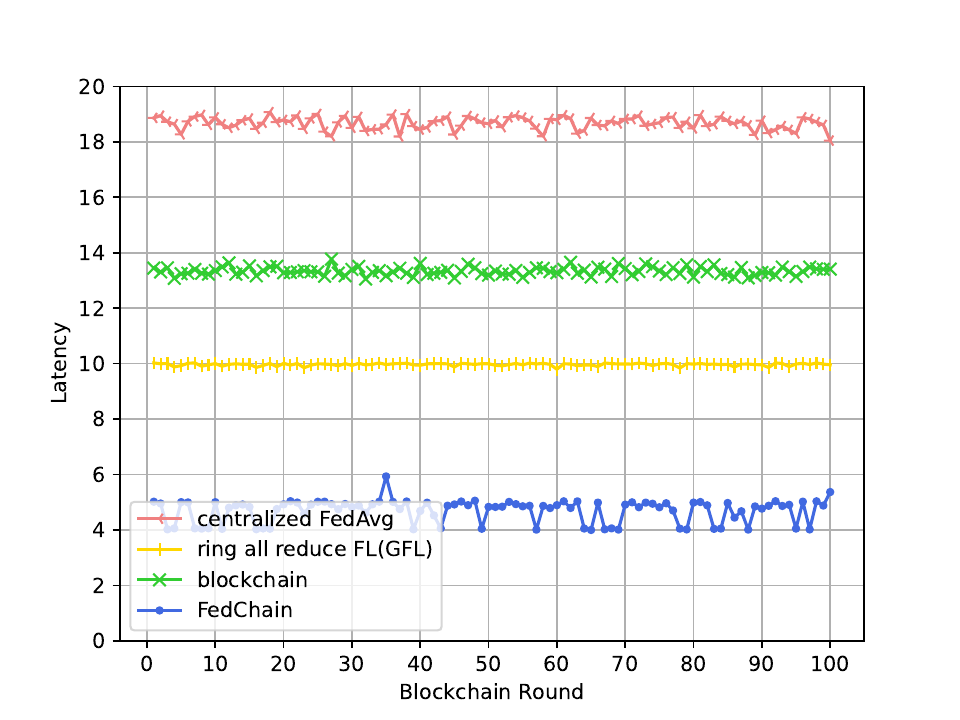}}
  \\
  \caption{Latency to complete proof protocol in different frameworks.}
  \label{fig:exp_latency}
  \vspace{0.2in}
\end{figure*}

\section{Evaluation}

To ensure the efficiency and federated learning model performance of our proposed FedChain consensus protocol, we evaluate FedChain with several schemes in protocol latency and federated learning model performance.

\subsection{Experiment Setup}




\subsubsection{Evaluated Datasets}
We evaluate our framework on 2 public datasets:
\begin{itemize}
    \item Mnist \cite{cohen2017emnist}: The MNIST database of handwritten digits, available from this page, has a training set of 60,000 examples, and a test set of 10,000 examples. It is a subset of a larger set available from NIST. The digits have been size-normalized and centered in a fixed-size image.
    \item Fashion-Mnist \cite{xiao2017/online}: Fashion-MNIST is a dataset of Zalando's article images—consisting of a training set of 60,000 examples and a test set of 10,000 examples. Each example is a 28x28 grayscale image, associated with a label from 10 classes. We intend Fashion-MNIST to serve as a direct drop-in replacement for the original MNIST dataset for benchmarking machine learning algorithms. It shares the same image size and structure of training and testing splits.
\end{itemize}

\subsubsection{Evaluated Schemes in Efficiency} We evaluate several schemes to compare their efficiency with our FedChain consensus protocol.

\begin{itemize}
    \item \textbf{FedAvg-based blockchain} \cite{bonawitz2019towards}: We consider directly introducing traditional FedAvg as the consensus protocol for the blockchain. In the FedAvg-based blockchain, we allow all the nodes within the blockchain to participate in centralized federated learning.
    \item \textbf{PoW-based blockchain}: We also take blockchain using traditional PoW as the consensus protocol to compare the latency between PoW and FedChain.
    \item \textbf{GFL (Ring all reduce FL)-based blockchain} \cite{hu2020gfl}: To prove our design of mining pool aggregation on decentralized federated learning performs better than other decentralized federated learning without mining pool aggregation, we take GFL, a ring all reduce communication-based decentralized federated learning as the consensus protocol.
\end{itemize}

\subsubsection{Evaluated Schemes in Model Performance} We evaluate several schemes to compare their model performance with our FedChain.

\begin{itemize}
    \item \textbf{FedAvg \cite{bonawitz2019towards}}: We directly take traditional federated learning architecture as one of the benchmarks. In the FedAvg-based consensus protocol, we allow all the nodes within the blockchain to participate in centralized federated learning.
    \item \textbf{FedProx \cite{li2020federated}}: In FedProx, Li et al. proposed adding a proximal term to the objective that helps to improve the stability of the method. This term provides a principled way for the server to account for heterogeneity associated with partial information.
\end{itemize}

\subsubsection{Evaluated Metrics} We evaluate our proposed FedChain scheme with other schemes in the following 2 metrics:

\begin{itemize}
    \item \textbf{Model Accuracy}: Since our applied mechanisms will affect the model accuracy, we need to consider whether it will have a bad effect on it.
    \item \textbf{Consensus Protocol Latency}: Since blockchain needs to execute effectively, we need to consider the latency of the whole system. We set the accuracy boundary to be 90\% which means the FL process will stop when the accuracy of FL reaches 90\%.
\end{itemize}

\subsection{Experiment Results}

\subsubsection{Latency Comparison}

\begin{figure}[htb!]
\centering
\includegraphics[width=0.5\textwidth]{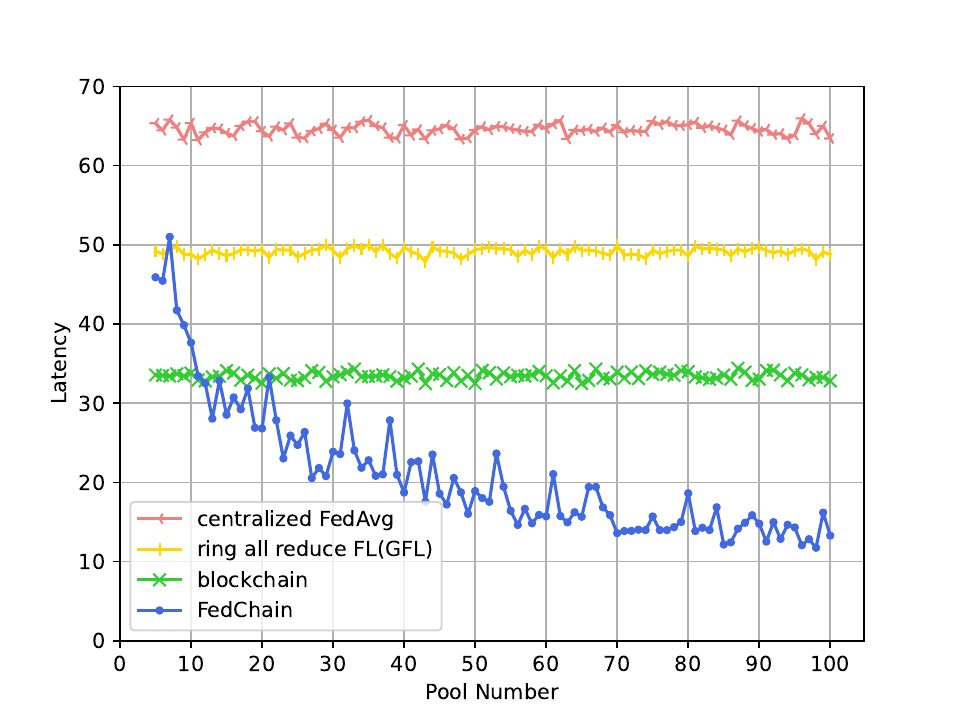}
\caption{Latency changes with different pool numbers.}
\label{fig:pool_num}
\end{figure}

\begin{figure}[htb!]
\centering
\includegraphics[width=0.5\textwidth]{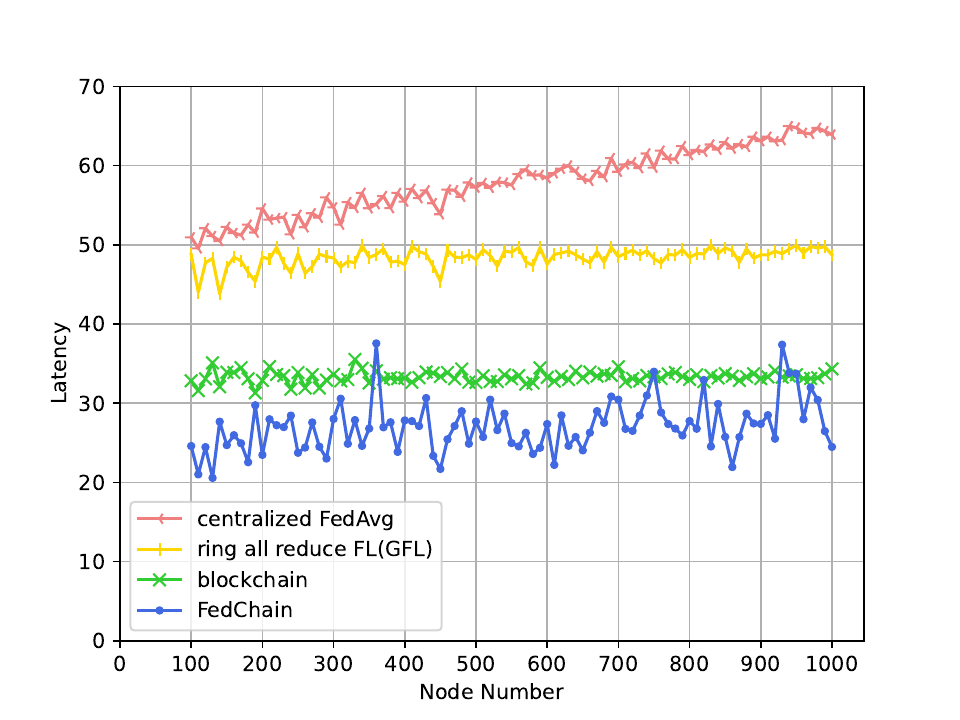}
\caption{Latency changes with different node numbers.}
\label{fig:node_num}
\end{figure}

\begin{figure*}[!htp]
  
  \centering
    \subfloat[MNIST \& non-iid=0.1]{\includegraphics[width=0.25\textwidth]{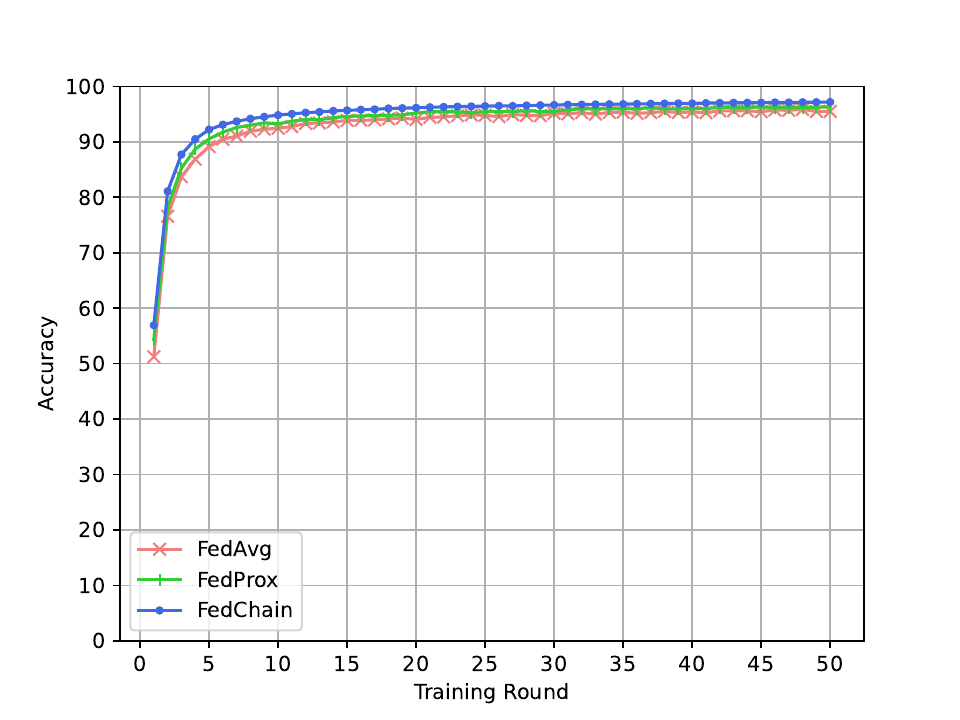}}
    \subfloat[MNIST \& non-iid=0.2]{\includegraphics[width=0.25\textwidth]{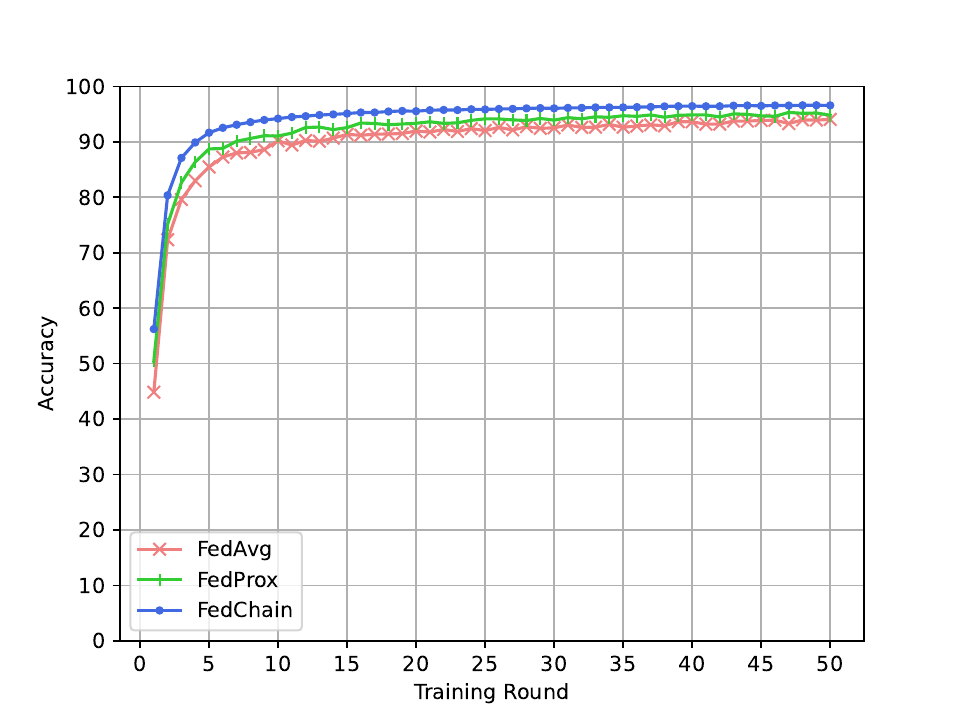}}
    \subfloat[MNIST \& non-iid=0.3]{\includegraphics[width=0.25\textwidth]{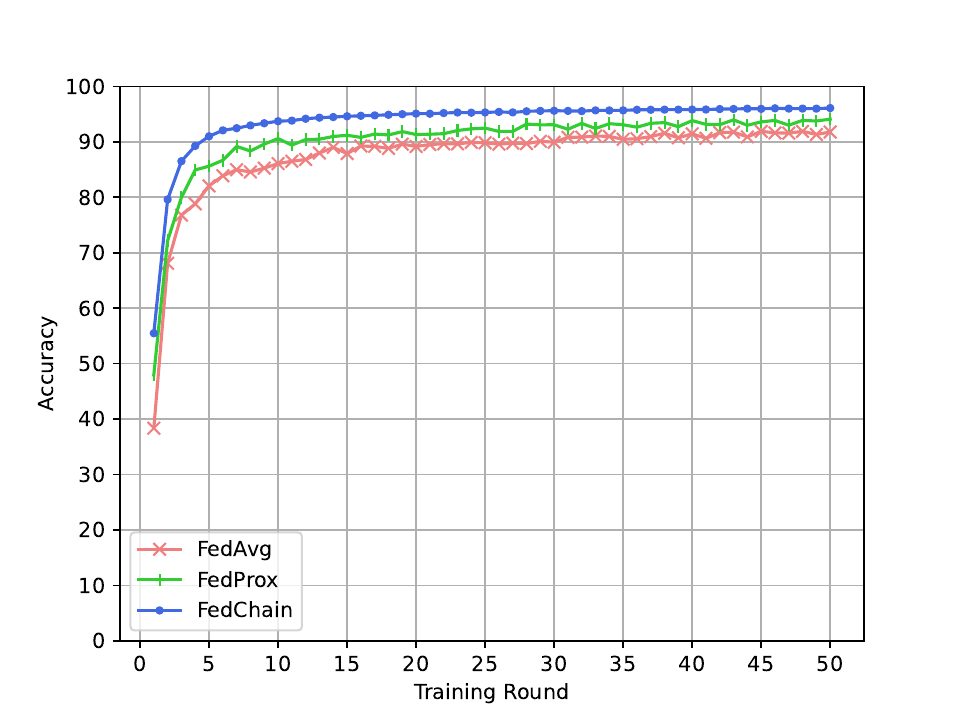}}
    \subfloat[MNIST \& non-iid=0.4]{\includegraphics[width=0.25\textwidth]{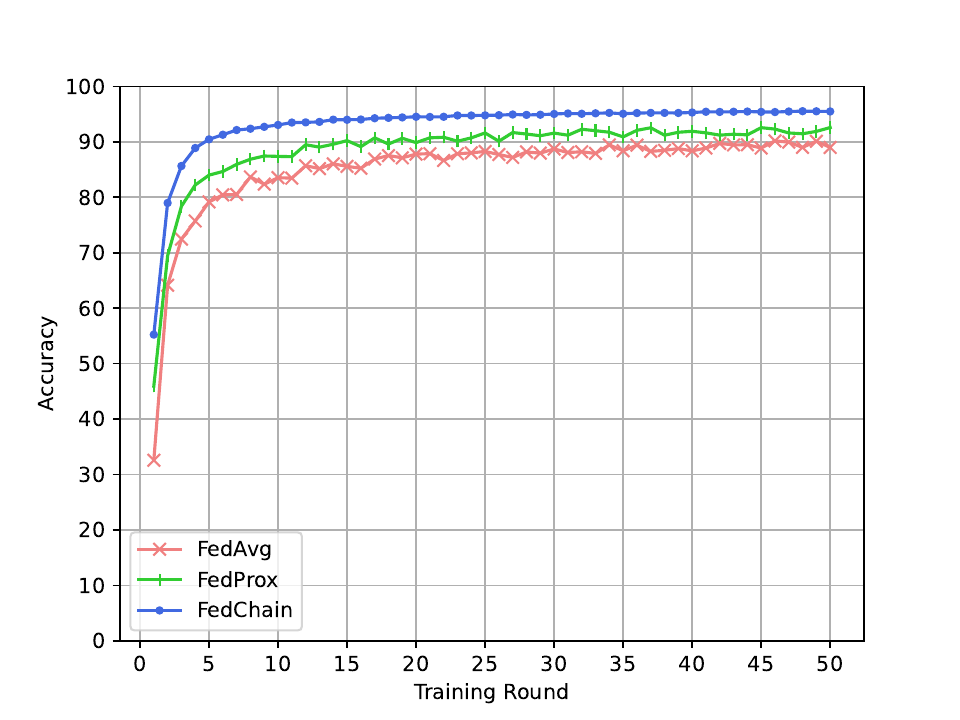}}
  \\
    \subfloat[MNIST \& non-iid=0.5]{\includegraphics[width=0.25\textwidth]{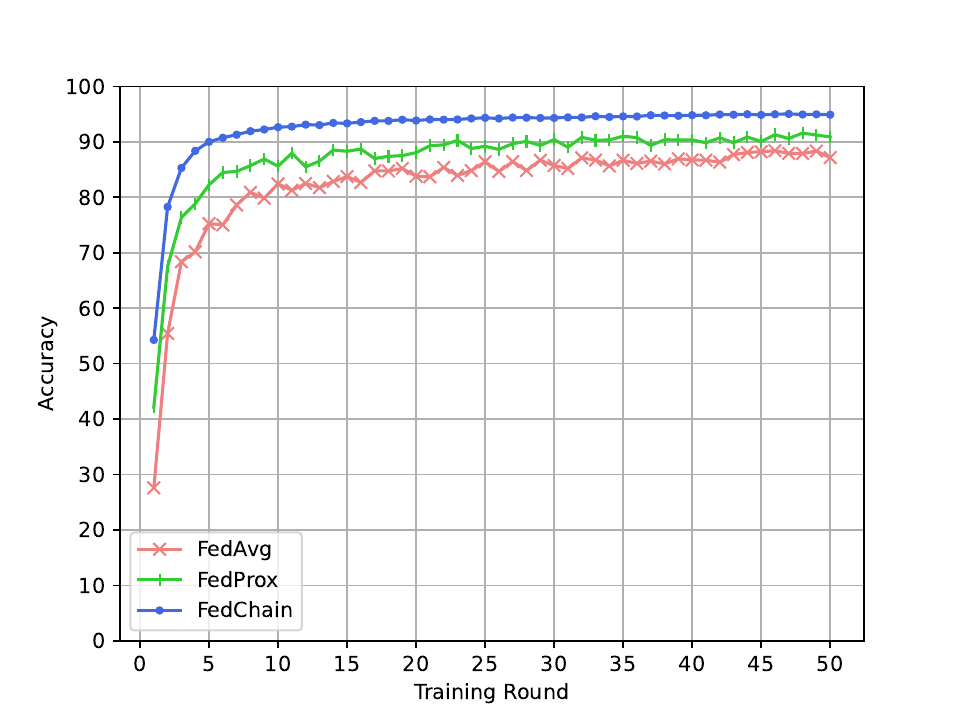}}
    \subfloat[MNIST \& non-iid=0.6]{\includegraphics[width=0.25\textwidth]{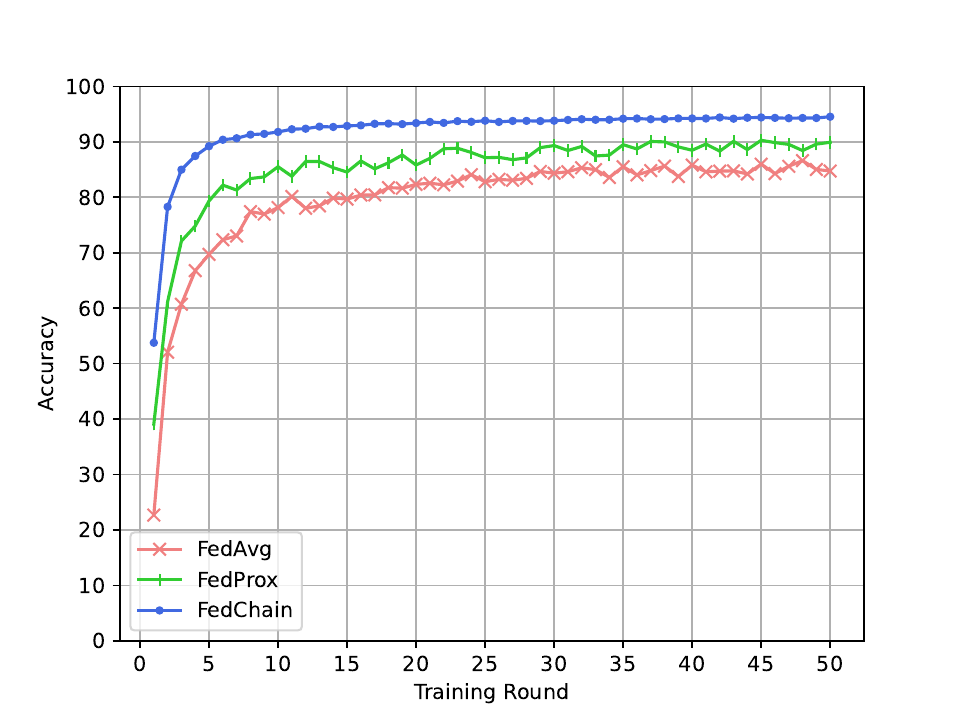}}
    \subfloat[MNIST \& non-iid=0.7]{\includegraphics[width=0.25\textwidth]{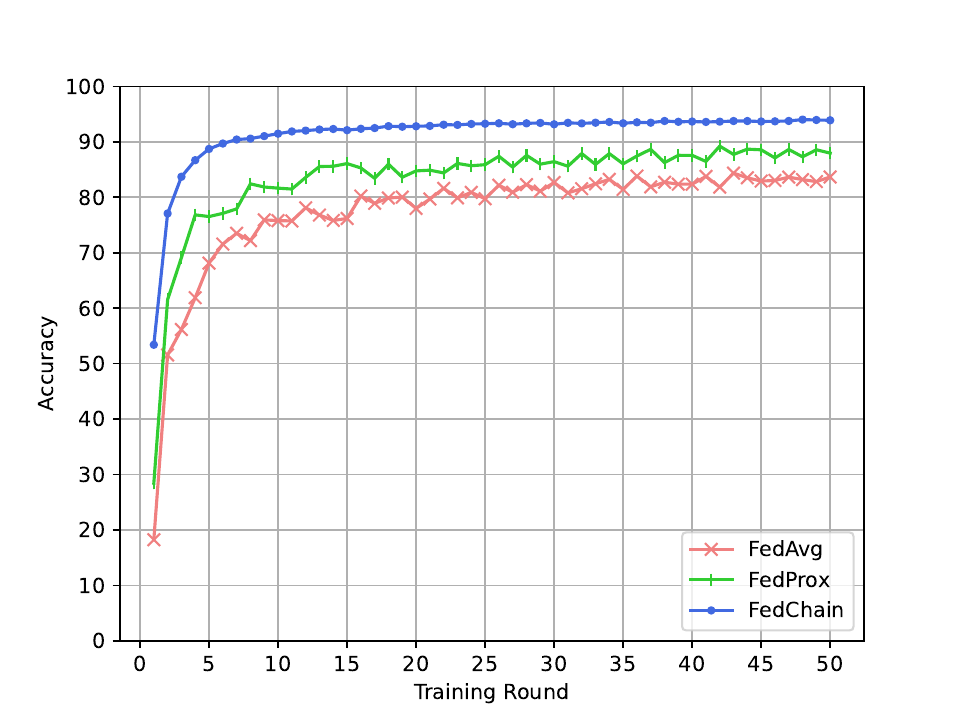}}
    \subfloat[MNIST \& non-iid=0.8]{\includegraphics[width=0.25\textwidth]{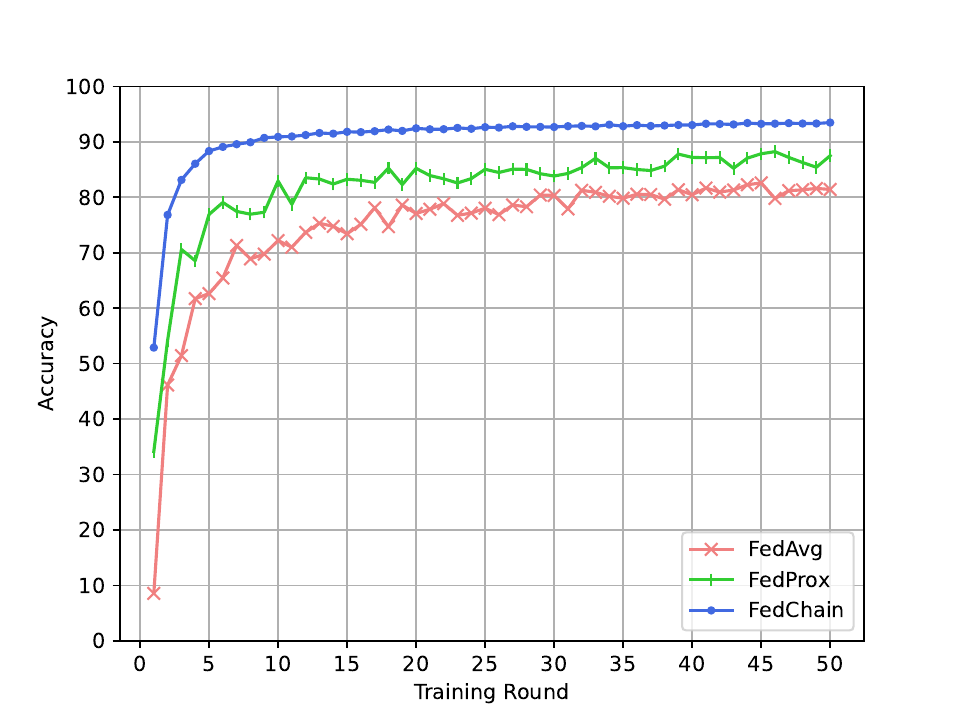}}
  \\
  \subfloat[Fashion-MNIST \& non-iid=0.1]{\includegraphics[width=0.25\textwidth]{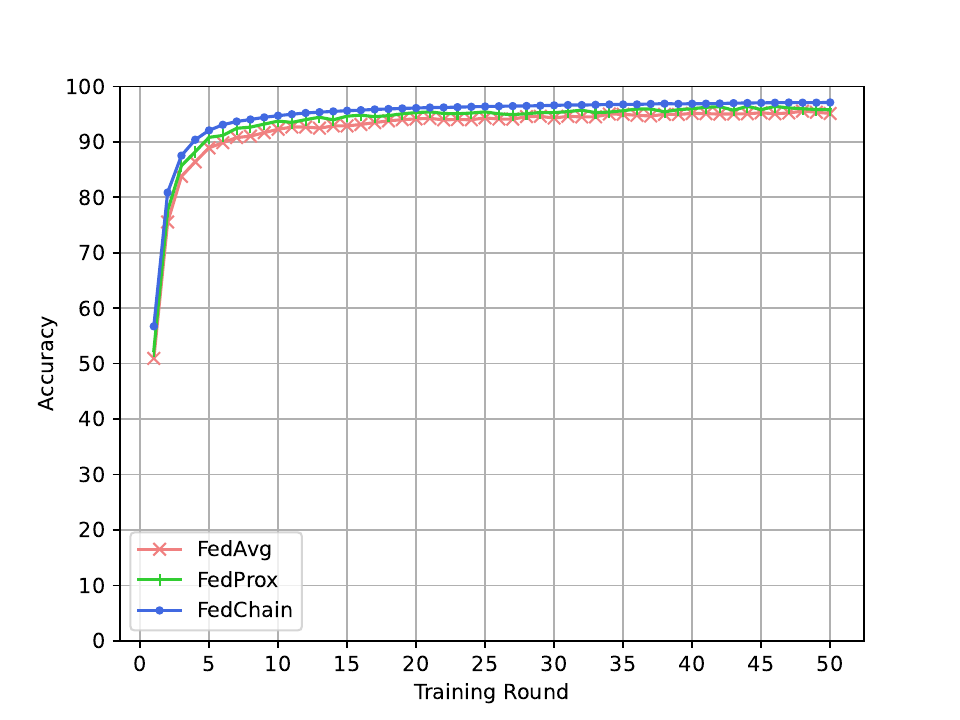}}
    \subfloat[Fashion-MNIST \& non-iid=0.2]{\includegraphics[width=0.25\textwidth]{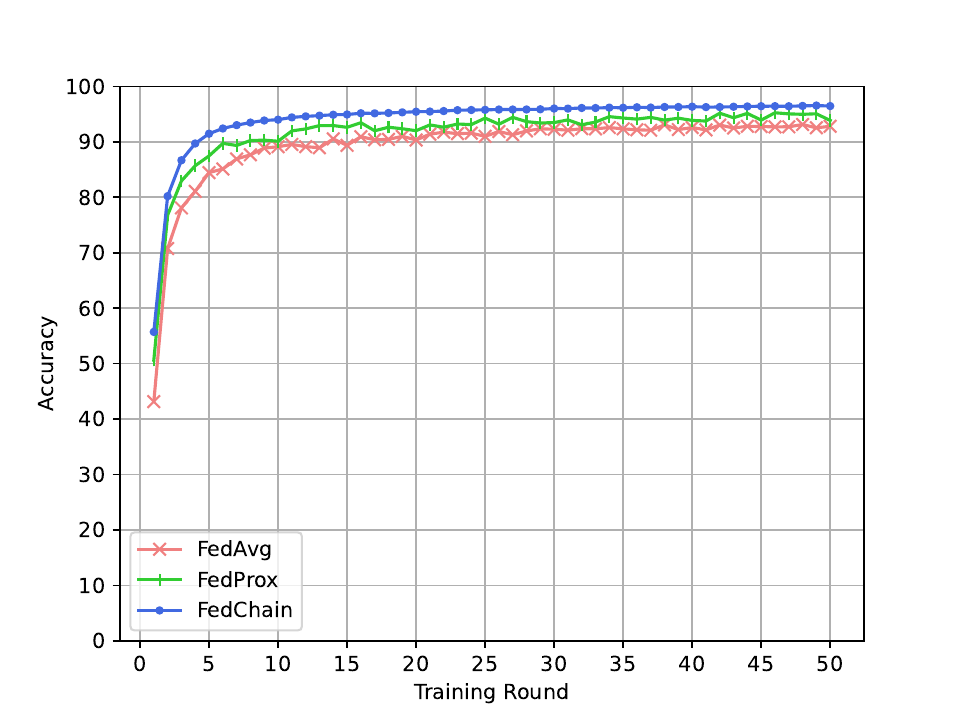}}
    \subfloat[Fashion-MNIST \& non-iid=0.3]{\includegraphics[width=0.25\textwidth]{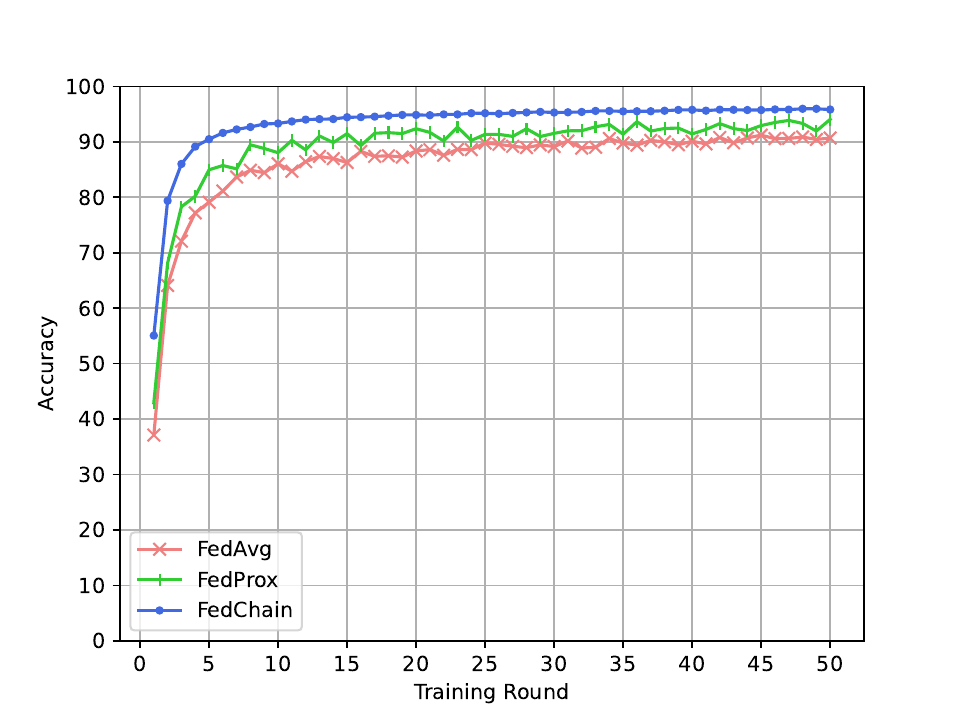}}
    \subfloat[Fashion-MNIST \& non-iid=0.4]{\includegraphics[width=0.25\textwidth]{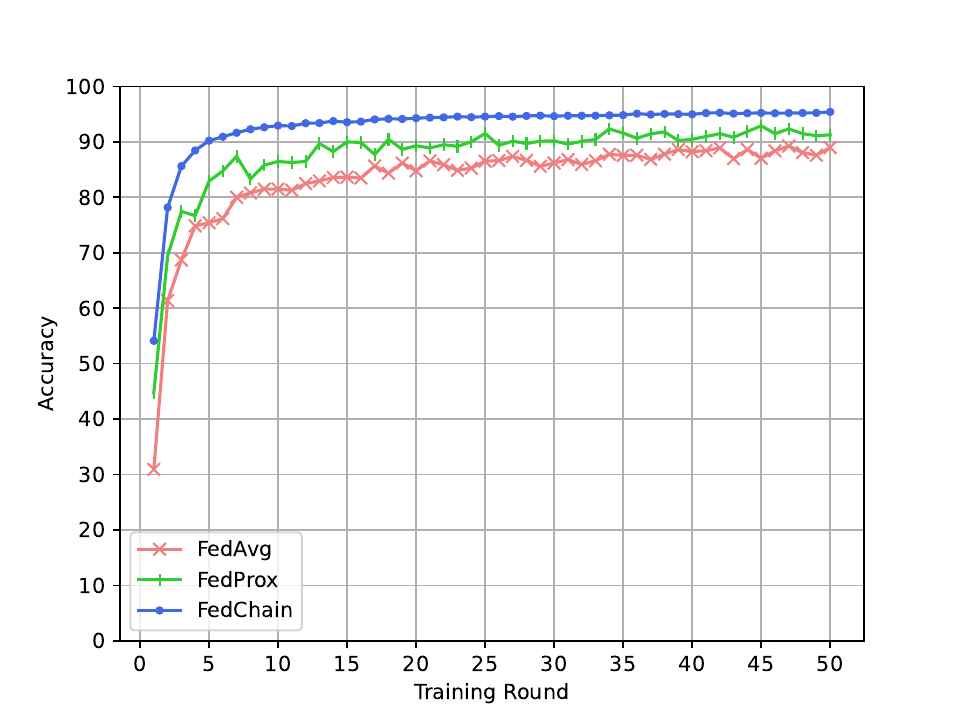}}
  \\
    \subfloat[Fashion-MNIST \& non-iid=0.5]{\includegraphics[width=0.25\textwidth]{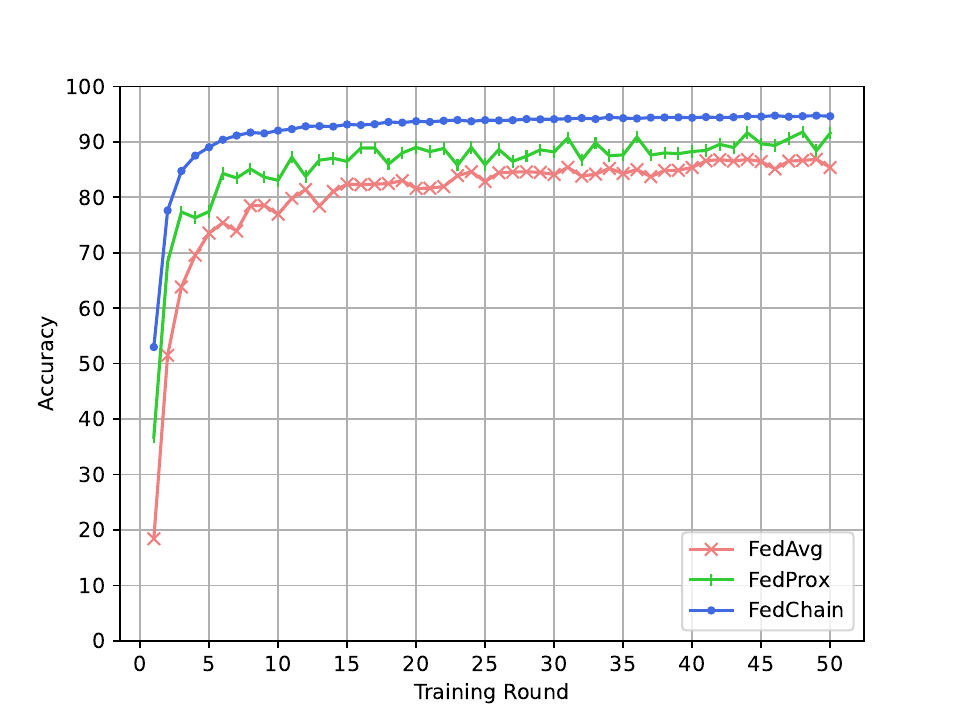}}
    \subfloat[Fashion-MNIST \& non-iid=0.6]{\includegraphics[width=0.25\textwidth]{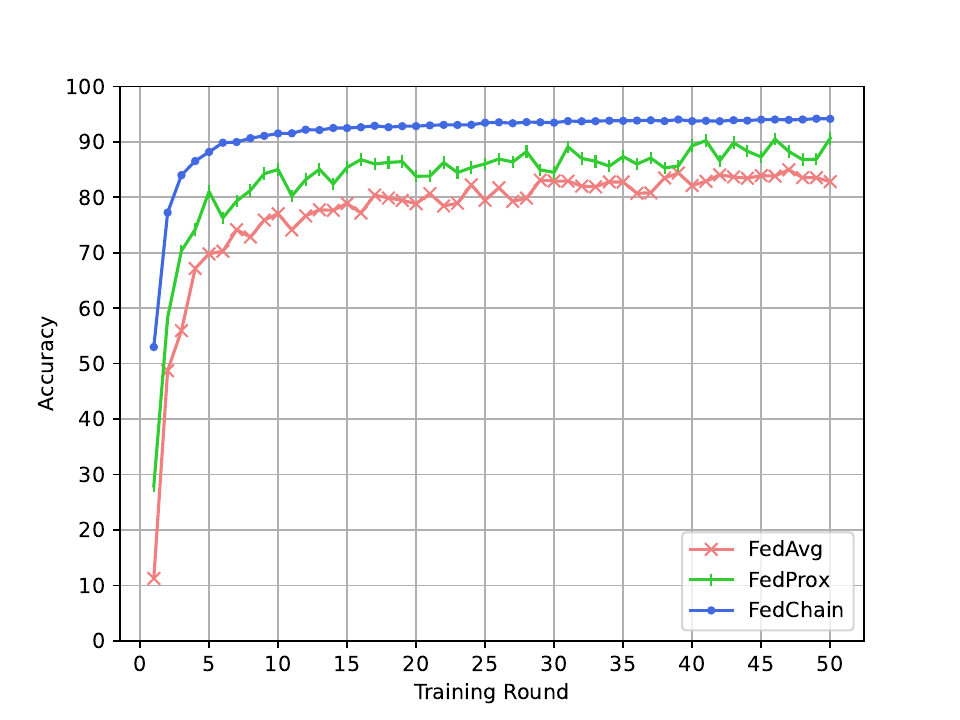}}
    \subfloat[Fashion-MNIST \& non-iid=0.7]{\includegraphics[width=0.25\textwidth]{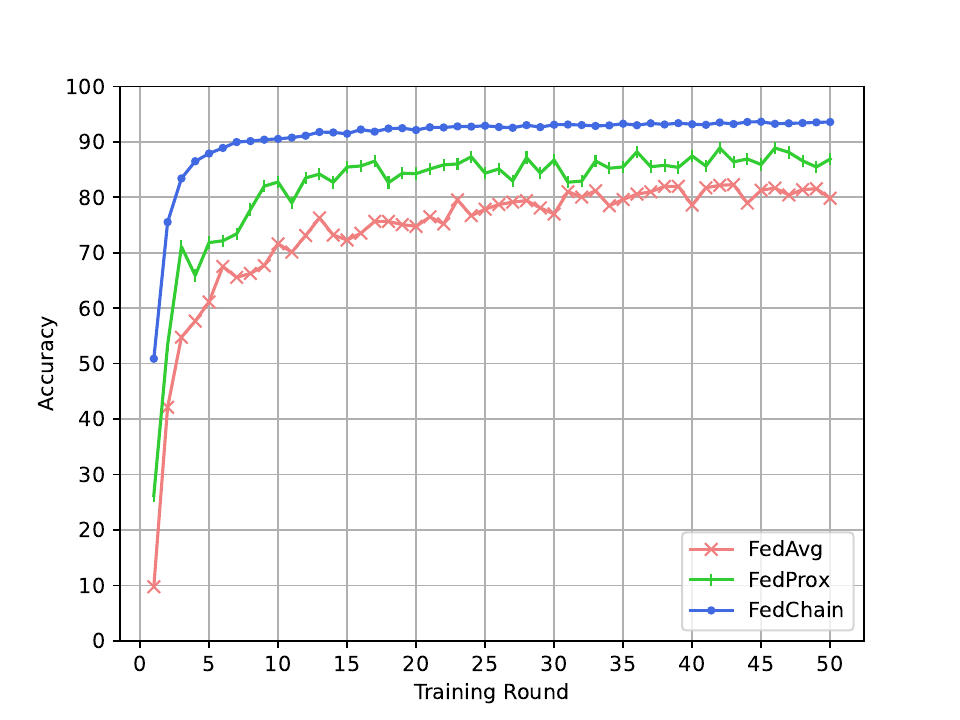}}
    \subfloat[Fashion-MNIST \& non-iid=0.8]{\includegraphics[width=0.25\textwidth]{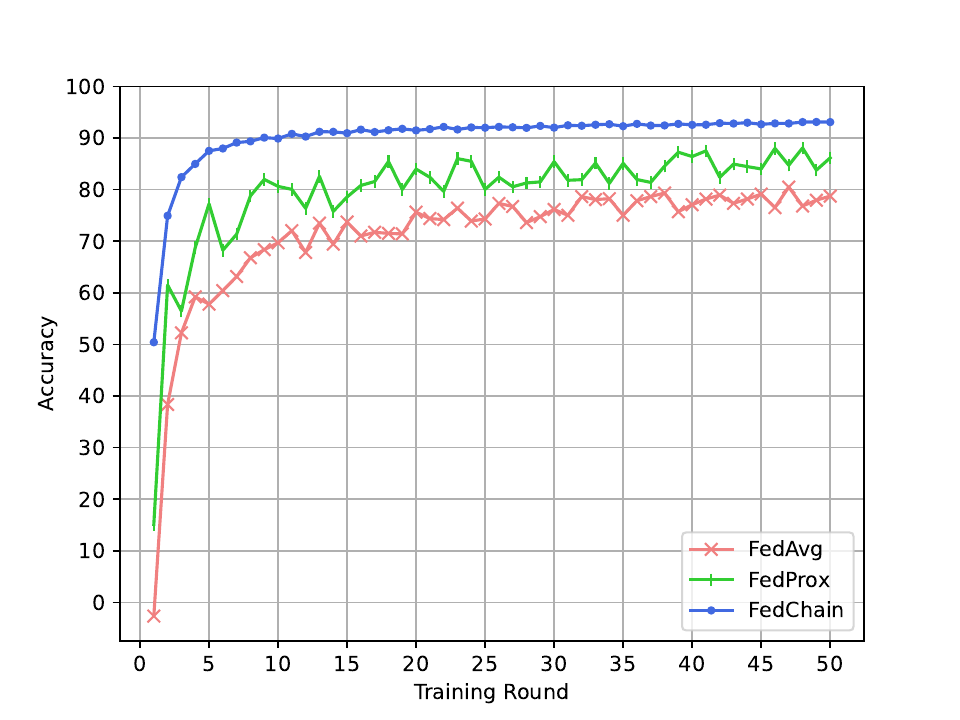}}
  \\
  \caption{Accuracy Comparison.}
  \label{fig:exp_accuracy}
  \vspace{0.2in}
\end{figure*}

As shown in Fig. \ref{fig:exp_latency}, we evaluate the centralized federated learning, decentralized federated learning, centralized federated learning with mining pool aggregation, and decentralized federated learning with mining pool aggregation (FedChain) as the consensus protocol for blockchain in our experiment. The experiment results show that the former 2 methods without mining pool aggregation have strong drawbacks compared with the latter 2 methods. The mining pool aggregation strongly reduces the latency for the consensus protocol to finish the proof for a mining block. Furthermore, FedChain takes decentralized federated learning instead of centralized federated learning, though rising some small latency but also bringing high-security support as shown in Fig. \ref{fig:exp_latency}.

\par Furthermore, we compare the effects of increasing node numbers. As shown in Fig. \ref{fig:node_num}, the increasing node number will increase the latency time for FedAvg to complete training. The key insight is that the centralized paradigm applied by FedAvg will cause the bandwidth of the central server to be the bottleneck for the whole system, thus increasing the latency for FedAvg to complete training. However, the rising node number will not have a large impact on the other 3 schemes (ring all reduce FL, blockchain, and FedChain) since all of them apply the decentralized paradigm.

\par The impact brought by pool number is also taken into account by us. As shown in Fig. \ref{fig:pool_num}, the latency of FedChain decreases steadily as the mining pool number increases. This indicates that the rising mining pool number will bring down the number of nodes within each mining pool, thus decreasing the latency.

\subsubsection{Model Performance Comparison}

The aim of the mining pool aggregation on the blockchain is to reduce the latency of federated learning to enable fast blockchain transactions. However, the reduction of the number of nodes participating in each mining pool's federated learning process will indeed reduce the model performance of the federated learning model (accuracy e.g). To confirm how the mining pool aggregation impacts federated learning performance, we evaluate federated learning without mining pool aggregation and federated learning with mining pool aggregation on the blockchain. As shown in Fig. \ref{fig:exp_accuracy}, the mining pool aggregation will not have a bad impact on the model performance. Furthermore, we compare our proposed FedChain scheme with 2 benchmarks: FedAvg and FedProx. The experiment result shows that when the data distribution among all the participants tends to be balanced (non-iid=0.1), the performance of FedChain does not have a strong difference with FedAvg and FedProx. While when the data distribution among all the participants tends to be imbalanced (non-iid>0.1), FedChain outperforms FedAvg and FedProx strongly. This is because FedChain's data distribution-based model optimization can handle the situation of non-iid data distribution in federated learning well thus making FedChain performs better than the other two algorithms.

\section{Conclusion}
The decentralized concept of blockchain has been rapidly adopted for various applications in the zero-trust environment, including the federated learning system. The energy waste resulting from the PoW implementation in blockchain has been a long-standing concern. In this paper, we presented FedChain, an efficient and secure consensus protocol based on proof of federated learning for blockchain, to the energy waste issue. We designed an efficient and secure federated learning task instead of a meaningless hash puzzle in the consensus protocol, implemented a node mining pool aggregation technique to accelerate the process, a secret sharing-based ring all-ring reduce to preserve each participant's privacy, a data distribution-based federated learning model optimization and a zero-knowledge proof-based federated learning model verification to preserve privacy. Our extensive experiment results demonstrated the effectiveness of our proposed consensus protocol, and analysis showed its security and privacy. In a word, FedChain provides a promising solution to motivate participants to join federated learning in the blockchain while avoiding significant energy waste.

\par Our future work mainly focuses on improving the efficiency of FedChain and testing FedChain on a bigger scale of implementation.

\bibliographystyle{ACM-Reference-Format}
\bibliography{sample-base}


\end{document}